\documentclass[twocolumn,showpacs,amsmath,amssymb,prb]{revtex4}

\usepackage{epsfig}
\usepackage{graphicx,color}

\newcommand{\eref}[1]{(\ref{#1})}

\begin{document}
\title{Effects of voltage fluctuations on the current correlations in
mesoscopic Y-shaped conductors}
\author{Shin-Tza Wu$^1$ and S.-K. Yip$^2$}
\affiliation{
$^1$Department of Physics, National Chung-Cheng University,
Chiayi 621, Taiwan\\
$^2$Institute of Physics, Academia Sinica, Nankang, Taipei 115,
Taiwan}
\date{February 22, 2006}

\begin{abstract}
We study current fluctuations in a phase coherent Y-shaped conductor
connected to external leads and voltage probes. The voltage probes
are taken to have finite impedances and thus can cause voltage
fluctuations in the circuit. Applying the Keldysh formulation and a
saddle point approximation appropriate for slow fluctuations, we
examine at zero temperature the feedback effects on the current
fluctuations due to the fluctuating voltages. We consider mesoscopic
Y-shaped conductors made of tunnel junctions and of diffusive wires.
Unlike two-terminal conductors, we find that for the Y-shaped
conductors the current moments in the presence of external
impedances cannot be obtained from simple rescaling of the bare
moments already in the second moments. As a direct consequence, we
find that the cross correlation between the output terminals can
become {\em positive} due to the impedances in the circuit. We
provide formulas for the range of parameters that can cause positive
cross correlations.
\end{abstract}
\maketitle
%###################
\section{Introduction}
%=====================
Current fluctuations in mesoscopic systems are of fundamental
interest theoretically and experimentally since they can reveal
information inaccessible from conductance measurements (see, for
example, Refs.~\onlinecite{BB}--\onlinecite{Nazarov_rev}). For
instance, non-equilibrium noises (or shot noises) can be used to
determine the effective charge of quasiparticles in transport; a
renowned example is the measurement of fractional charges in quantum
Hall systems.\cite{FQHE} On the other hand, the statistics of the
charge carriers can be studied in multiterminal setups. The quantity
of interest in this case is the cross correlation between different
terminals.\cite{BB,BS} A typical experiment of this type is the
solid-state analogue of the Hanbury Brown-Twiss experiment in
quantum optics;\cite{QO} the electron versions of such experiments
have shown elegantly that Fermi statistics suppresses current
fluctuations in comparison with uncorrelated charge
carriers.\cite{HBT1,HBT2} There have also been proposals for probing
the fractional statistics of quasiparticles in quantum Hall
systems\cite{Goldman} base on Hanbury Brown-Twiss type
setups.\cite{Safi,Vish,Fradkin}

Typically, in measurements of cross correlations, one
injects an incident beam of charge carriers and then splits
the beam into two parts using a ``beam splitter", such as a
Y-shaped conductor.\cite{HBT1,HBT2} By measuring the
intensity correlation between the output beams, one can
extract information regarding the cross correlations. In
most situations, current measurements involve coupling of
the sample to external circuits. If the measurement
circuit can be idealized as having zero impedance,\cite{HBT2}
the voltage across the sample would be non-fluctuating and
the current fluctuations are entirely due to intrinsic
properties (such as statistics, for example) of the carriers.
If, however, the measuring circuit has non-negligible
impedance,\cite{HBT1} the voltage across the sample then
becomes fluctuating and the current fluctuations in this
case will be modified due to the voltage fluctuations. This
is the feedback effect we will be investigating in this paper.

\begin{figure}
\includegraphics*[width=50mm]{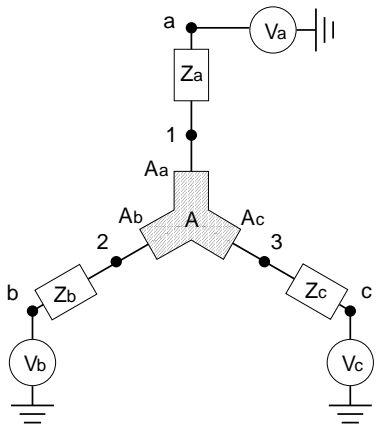}
\caption{\small Schematic for the system considered in this paper.
The arms $A_a$, $A_b$, $A_c$ of a mesoscopic Y-shaped conductor $A$
are connected to external leads biased, respectively, at voltages
$V_a$, $V_b$, and $V_c$. The leads are assumed to have impedances
$Z_a$, $Z_b$, and $Z_c$, which are schematized as external resistors
connected to the sample arms. The nodes $1,2,3$ between the sample
arms and the resistors are where voltage fluctuations set in.}
\label{circuit}
\end{figure}

Feedback effects on current moments have previously been considered
for two-terminal cases at zero temperature and at finite
temperatures.\cite{BB,RSP,KNB,Kind,BKN} The results based on the
Langevin formalism concluded that the second moments of current
fluctuations can be obtained from the corresponding zero-impedance
(intrinsic or ``bare") values by simple scaling. However,it was
shown, using a Keldysh technique\cite{KNB,Kind} and the Langevin
formalism,\cite{Kind,BKN} that this rescaling breaks down at the
third moment. In this work, we study a three-terminal setup (see
Fig.~\ref{circuit}) using the Keldysh formulation, which complements
a previous analysis by the present authors using the Langevin
formulation.\cite{WY05} Although the effects of external impedances
on current fluctuations in multi-terminal circuits have also been
considered by B\"uttiker and coworkers,\cite{BB} they focus mainly
on a multiprobe measurement of a two-terminal conductor and thus not
directly our geometry here (see, however, Ref.~\onlinecite{RB} that
we shall refer to later).

We will study a mesoscopic, phase coherent Y-shaped conductor
(sample $A$ in Fig.~\ref{circuit}) connected to a measuring circuit
with finite impedances (schematized as $Z_a$, $Z_b$, $Z_c$ in
Fig.~\ref{circuit}). The sample arms of $A$ will be taken to be
either tunnel junctions or diffusive wires. In the absence of the
external impedances, it is well known that the cross correlation
between different terminals, say $b$ and $c$, is negative due to
Fermi statistics of the electrons.\cite{FS} With finite impedances
in the measuring circuit, however, we will find that voltage
fluctuations can modify current fluctuations significantly. In
particular, the second moment cannot be obtained from simple
rescaling of the corresponding zero-impedance (or ``bare")
expressions. For instance, the cross correlation will acquire
contributions from noise correlators. Since the bare noise
correlators are always positive, it is then possible to have {\em
positive} cross correlations in the appropriate parameter
regime.\cite{WY05}

Theoretical predictions for positive cross correlations in
multiterminal setups had previously been made for systems with
terminals of non-Fermi liquid ground states,\cite{Buttiker_rev} such
as superconductors,\cite{SC} quantum Hall states,\cite{Texier,Safi}
Luttinger liquids,\cite{Safi,Crepieux} and
ferromagnets.\cite{Taddei,CBB} In normal systems it has also been
predicted for systems with capacitively coupled contacts.\cite{MB00}
Our finding of positive cross correlations due to feedback effects
thus provides a new mechanism for positive cross correlation in
normal systems. It is of direct experimental relevance since
feedback effects are of crucial importance in almost all experiments
(cf., for example, Ref.~\onlinecite{RSP}). A recent preprint by
Rychkov and B\"uttiker \cite{RB} studies the sign change of current
cross correlations in three-terminal conductors due to inelastic
scattering, and hence voltage fluctuations. Whereas our voltage
fluctuations occur outside our coherent conductor (at nodes 1--3 in
Fig.~1), they considered the situation where inelastic scattering
occurs inside the Y-shaped conductor where the arms joint together.
In our case, as we shall see, the sign of the cross correlation
crucially depends on {\it how} the external impedances are placed:
for $V_b=V_c$ (as we shall consider below), we find that $Z_b$,
$Z_c$ tend to make the cross correlation between terminals $b$ and
$c$ more positive, while $Z_a$ would make it more {\it negative}
[see Eq.~\eref{regions} and Appendix C].

We shall derive in Sec.~\ref{sec_formln} analytic formulas for the
current moments using the Keldysh technique. In
Sec.~\ref{sec_cross_corrln} we will then present numerical results
for the cross correlation and discuss in detail its sign change in
different parameter regimes. To help focus on the main points, we
relegate most details to the Appendices. Finally in
Sec.~\ref{sec_conclu} we summarize and further discuss our results.

\section{Formulation}
\label{sec_formln}
%====================
We consider a system schematized as in Fig.~\ref{circuit}, where a
phase coherent mesoscopic Y-shaped conductor $A$ (the ``sample")
with arms $A_a$, $A_b$, and $A_c$ is connected to a measuring
circuit. Each arm $A_\alpha$ ($\alpha=a,b,c$) is connected to an
external lead $\alpha$ biased at voltage $V_\alpha$. The arm
$A_\alpha$ is taken to have conductance $G_\alpha$ and the lead
connected to it has impedance $Z_\alpha$. For convenience, we define
the dimensionless quantities $g \equiv (h/e^2)(G_a+G_b+G_c)$,
$\eta_\alpha\equiv G_\alpha/(G_a+G_b+G_c)$, and $z_\alpha\equiv
(e^2/h)Z_\alpha$. Note that it follows from these definitions
$\eta_a+\eta_b+\eta_c=1$. Without lost of generality, we shall take
$V_a\geq V_b\geq V_c$ and the downward direction as the positive
current direction.

To take voltage fluctuations into account, we apply the Keldysh
technique developed in Ref.~\onlinecite{KNB}. In this formulation,
the current moments are obtained from the generating functional
\begin{widetext}
\begin{eqnarray}
{\cal Z}[\Phi,\chi] = \left\langle \overleftarrow{T}
\exp\left\{\frac{i}{e} \int d t\left[ \Phi(t)+\frac{1}{2}\chi(t)
\right] \hat{I}(t)\right\} \overrightarrow{T} \exp\left\{\frac{i}{e}
\int d t\left[-\Phi(t)+\frac{1}{2}\chi(t) \right] \hat{I}(t)\right\}
\right\rangle . \label{Z_def}
\end{eqnarray}
\end{widetext}
Here $e$ is the charge of an electron, $\hat{I}$ is the current
operator, $\overleftarrow{T}$ and $\overrightarrow{T}$ are,
respectively, the time ordering operators in ascending and
descending directions; $\Phi(t) = (e/\hbar)\int_0^t V(t') dt'$
stands for the accumulated phase, and $\chi(t)$ is the counting
field. Evaluation of the full expression will then allow one to
obtain the {\em full counting statistics}\cite{FCS} of transported
charges. At the same time, current moments of any order can also be
obtained from functional derivatives of $\ln{\cal Z}$ with
respective $\chi(t)/e$.\cite{KNB}

To implement the Keldysh approach to current fluctuations in the
presence of external impedances in our problem, therefore, our first
task is to set up the auxiliary fields $\Phi$ and $\chi$, taking
into account effects of the external resistors. At the leads
$\alpha=a,b,c$, we denote the counting fields as $\chi_\alpha$ and
the accumulated phases during the detection time $\tau$ as
$\Phi_\alpha=(e/\hbar)\int_0^\tau V_\alpha dt$. At the nodes
$k=1,2,3$, these variables take unknown values $\chi_k, \Phi_k$,
which have to be integrated over all possible values in the
generating functional for the total system. Generalizing the
expressions of Ref.~\onlinecite{KNB}, one can then obtain the
generating functional for the total system from the convolution of
the generating functionals of the leads (${\cal Z}_\alpha$) and the
sample (${\cal Z}_A$), namely
\begin{widetext}
\begin{eqnarray}
{\cal Z}_{tot}[\Phi_a,\Phi_b,\Phi_c,\chi_a,\chi_b,\chi_c]
&=&
\int \prod_{k=1}^3{\cal D}\Phi_k \,\, {\cal D}\chi_k \,\,
{\cal Z}_A[\Phi_1,\Phi_2,\Phi_3,\chi_1,\chi_2,\chi_3] \,\,
{\cal Z}_a[(\Phi_a-\Phi_1),(\chi_a-\chi_1)]
\nonumber \\ && \hspace*{25mm}
{\cal Z}_b[(\Phi_b-\Phi_2),(\chi_b-\chi_2)] \,\,
{\cal Z}_c[(\Phi_c-\Phi_3),(\chi_c-\chi_3)] \, .
\label{Z_tot}
\end{eqnarray}
\end{widetext}
Since we are interested in the zero temperature limit, as explained
in Ref.~\onlinecite{KNB}, for slow enough phase fluctuations ({\em
ie.,}~in the low frequency regime) we can approximate the generating
functional by defining the action $S$ so that
\begin{eqnarray}
{\cal Z}[\Phi(t),\chi(t)] =
\exp\left\{ \int d t S[\dot{\Phi}(t),\chi(t)]\right\} \, ,
\label{Z_saddle}
\end{eqnarray}
where $\dot{\Phi}=d\Phi/dt$. A saddle point approximation to ${\cal
Z}_{tot}$ then yields from Eqs.~\eref{Z_tot} and \eref{Z_saddle}
\begin{eqnarray}
&& \ln {\cal Z}_{tot} =
\tau S_{tot}(
\dot{\Phi}_a, \dot{\Phi}_b, \dot{\Phi}_c,
\chi_a, \chi_b, \chi_c)
\nonumber \\
&=& \tau S_A(\dot{\Phi}_1,\dot{\Phi}_2,\dot{\Phi}_3,
\chi_1,\chi_2,\chi_3)
+ \tau S_a(\dot{\Phi}_a-\dot{\Phi}_1,\chi_a-\chi_1)
\nonumber \\
&+& \tau S_b(\dot{\Phi}_b-\dot{\Phi}_2,\chi_b-\chi_2)
+ \tau S_c(\dot{\Phi}_c-\dot{\Phi}_3,\chi_c-\chi_3) \, ,
\end{eqnarray}
where $\tau$ is the detection time and
$\dot{\Phi}_k$, $\chi_k$ ($k=1,2,3$) are evaluated at
their saddle point values. One can thus obtain current
moments of any order from derivatives of $\tau S_{tot}$
with respect to the counting fields $\chi_\alpha$.\cite{KNB}

To carry out the calculation explicitly, we will need expressions
for the actions in the above formula. For the leads, there are exact
expressions for the actions (see below). However, in general, this
is not the case for the action $S_A$ for the sample. As we will be
interested in the second-order current moments in the
zero-temperature regime, an expression linear in $\Phi_\alpha$ and
second order in $\chi_\alpha$ will suffice. Therefore, the next step
of our calculation is to expand the total action to linear order in
$\Phi_\alpha$ and quadratic order in $\chi_\alpha$. The coefficients
will then yield information for the {\em renormalized} current
moments (that is, modified current moments due to the impedances in
the circuit -- see below), as we shall now show.

For the leads, the Keldysh actions can be found exactly.
The lead $\alpha$ that is connected to the arm $A_\alpha$
via node $k$ has the action
\begin{eqnarray}
S_\alpha =
\frac{i(\chi_\alpha-\chi_k)(\dot{\Phi}_\alpha-\dot{\Phi}_k)}
{2\pi z_\alpha}\, .
\end{eqnarray}
Alternatively, defining $\sigma\equiv i\chi$ and
$\phi\equiv \Phi/(2\pi)$, we have
\begin{eqnarray}
\tau S_\alpha = \frac{(\sigma_\alpha-\sigma_k)(\phi_\alpha-\phi_k)}
{z_\alpha} \, .
\end{eqnarray}
The total action is thus
\begin{eqnarray}
\tau S_{tot} &=&
\frac{(\sigma_a-\sigma_1)(\phi_a-\phi_1)}{z_a}+
\frac{(\sigma_b-\sigma_2)(\phi_b-\phi_2)}{z_b}
\nonumber \\ && +
\frac{(\sigma_c-\sigma_3)(\phi_c-\phi_3)}{z_c} + \tau S_A \, .
\label{S_tot}
\end{eqnarray}
To linear order in the phase variables $\phi_k$, the action
for the sample $A$ can be expressed
\begin{eqnarray}
\tau S_A =
{\cal S}^{(b)}(\phi_1-\phi_2) +
{\cal S}^{(c)}(\phi_1-\phi_3) \, , \label{S_A}
\end{eqnarray}
where ${\cal S}^{(\alpha)}$ are functions of $\sigma_1$, $\sigma_2$,
and $\sigma_3$. Expansions of the ${\cal S}^{(\alpha)}$'s then
generate quantities proportional to the current moments.\cite{KNB}
To second order we have
\begin{eqnarray}
{\cal S}^{(\alpha)} &\simeq&
s_{b}^{(\alpha)} (\sigma_1-\sigma_2) +
s_{c}^{(\alpha)} (\sigma_1-\sigma_3) +
\frac{s_{bb}^{(\alpha)}}{2!} (\sigma_1-\sigma_2)^2
\nonumber \\ &&
+ \frac{s_{cc}^{(\alpha)}}{2!} (\sigma_1-\sigma_3)^2
+ s_{bc}^{(\alpha)} (\sigma_1-\sigma_2)(\sigma_1-\sigma_3)
\, . \label{S_alpha_expand}
\end{eqnarray}
The expansion coefficients are related to the {\em bare} current
moments ({\em ie.}, current moments in the absence of external
resistors -- see below). For example, $s_{b}^{(b)}$ is the
dimensionless conductance ({\em ie}.~conductance in units of
$e^2/h$) of arm $A_b$ due to the potential difference between nodes
1 and 2. Likewise, $s_{bb}^{(c)}$ is proportional to the current
noise at arm $A_b$ due to the potential difference between nodes 1
and 3. Since there is no any external resistors between each pair of
nodes, the moments related to these coefficients are therefore the
``bare" ones. These can be calculated in many methods and the
results are summarized in Appendix A.

From the saddle point condition
\begin{eqnarray}
\frac{\partial}{\partial \phi_k}
\big( \tau S_{tot} \big) = 0
\, , \qquad (k=1,2,3) \, ,
\end{eqnarray}
we get
\begin{eqnarray}
&& -\frac{\sigma_a-\sigma_1}{z_a} +
{\cal S}^{(b)} + {\cal S}^{(c)} = 0
\, , \nonumber \\
&& -\frac{\sigma_b-\sigma_2}{z_b} -
{\cal S}^{(b)} = 0
\, , \nonumber \\
&& -\frac{\sigma_c-\sigma_3}{z_c} -
{\cal S}^{(c)} = 0 \, .
\label{saddle}
\end{eqnarray}
Using \eref{saddle} and \eref{S_A} in Eq.~\eref{S_tot}, we find that
at the saddle point the total action is reduced to
\begin{eqnarray}
\tau S_{tot} =
{\cal S}^{(b)} (\phi_a-\phi_b) +
{\cal S}^{(c)} (\phi_a-\phi_c) \, .
\label{S_tot_saddle1}
\end{eqnarray}
Note that we have eliminated all the intermediate variables
$\phi_k$; ${\cal S}^{(b)}$ and ${\cal S}^{(c)}$ depend only on the
intermediate variables $\sigma_k$ ($k=1,2,3$). Also, it is
interesting to note that \eref{S_tot_saddle1} takes the form of
$\tau S_A$ (see Eq.~\eref{S_A}) with $\phi_1=\phi_a$,
$\phi_2=\phi_b$, and $\phi_3=\phi_c$. However, this simplicity is
deceptive and one should not take it literally. One can check easily
that it would otherwise lead to contradictions.

While Eq.~\eref{S_tot_saddle1} is completely general, for
simplicity, we shall from now on consider the case of
$\phi_b=\phi_c$ by setting $V_a=V$ and $V_b=V_c=0$. Using the first
of the saddle point equations \eref{saddle}, one finds that the
total action becomes
\begin{eqnarray}
\tau S_{tot} =
\frac{(\sigma_a-\sigma_1)(\phi_a-\phi_b)}{z_a}\, .
\label{tau_S_tot}
\end{eqnarray}
Note that there is now only one unknown, $\sigma_1$, that remains.
As the total action contains contributions from the external
resistors, an expansion of $\tau S_{tot}$ with respect to
$\sigma_a$, $\sigma_b$, and $\sigma_c$ can yield the {\em
renormalized} current moments ({\em ie.}, current moments in the
presence of external impedances) of any orders. For this purpose, in
view of Eq.~\eref{tau_S_tot}, we need only expand $\sigma_1$ in
terms of $\sigma_a$, $\sigma_b$, and $\sigma_c$. As we shall see,
interesting effects arise already in the second moments. Thus, we
expand $\sigma_1$ to second order and get
\begin{eqnarray}
\frac{(\sigma_a-\sigma_1)}{z_a} &\simeq&
C_{b} (\sigma_a-\sigma_b) +
C_{c} (\sigma_a-\sigma_c)
\nonumber \\ &&
+ \frac{C_{bb}}{2!} (\sigma_a-\sigma_b)^2
+ \frac{C_{cc}}{2!} (\sigma_a-\sigma_c)^2
\nonumber \\ &&
+ C_{bc} (\sigma_a-\sigma_b)(\sigma_a-\sigma_c) \, .
\label{C_coeff}
\end{eqnarray}
The expansion coefficients here are directly related to the
renormalized current moments: $(e^2/h)C_\alpha$ yields the
conductance and $(e^3V/h)C_{\alpha\beta}$ the noise correlator/cross
correlation. For brevity, we shall later on refer to them simply as
the conductance and the noise correlator/cross correlation (and
likewise for their bare counterparts).

It is not hard to solve the expansion coefficients in
Eq.~\eref{C_coeff} (see Appendix B). We find the first order
coefficients
\begin{eqnarray}
C_{b} =  \frac{1}{z_t}
\left[ (P+S) s_{b}^{(b)} + (Q+R) s_{b}^{(c)} \right] \, ,
\nonumber \\
C_{c} =  \frac{1}{z_t}
\left[ (P+S) s_{c}^{(b)} + (Q+R) s_{c}^{(c)} \right] \, ,
\label{C_1st}
\end{eqnarray}
and the second order coefficients
\begin{widetext}
\begin{eqnarray}
C_{bb} &=& \frac{1}{z_t^3}
\left\{
(P+S)
\left[P^2 s_{bb}^{(b)}+Q^2 s_{cc}^{(b)}+2PQ s_{bc}^{(b)}\right]
+ (Q+R)
\left[P^2 s_{bb}^{(c)}+Q^2 s_{cc}^{(c)}+2PQ s_{bc}^{(c)}\right]
\right\} \, ,
\label{Cbb} \\
C_{cc} &=& \frac{1}{z_t^3}
\left\{
(P+S)
\left[S^2 s_{bb}^{(b)}+R^2 s_{cc}^{(b)}+2SR s_{bc}^{(b)}\right]
+ (Q+R)
\left[S^2 s_{bb}^{(c)}+R^2 s_{cc}^{(c)}+2SR s_{bc}^{(c)}\right]
\right\} \, ,
\label{Ccc} \\
C_{bc} &=& \frac{1}{z_t^3}
\left\{
(P+S)
\left[PS s_{bb}^{(b)}+QR s_{cc}^{(b)}+(PR+QS) s_{bc}^{(b)}\right]
\right.\nonumber \\ &+& \left.
(Q+R) \left[PS s_{bb}^{(c)}+QR s_{cc}^{(c)}+(PR+QS)
s_{bc}^{(c)}\right] \right\} \, . \label{Cbc}
\end{eqnarray}
In these equations
\begin{eqnarray}
z_t &\equiv& g^2 \eta_a\eta_b\eta_c \left[
\left(z_a + \frac{1}{g\eta_a}\right)
\left(z_b + \frac{1}{g\eta_b}\right)
+\left(z_b + \frac{1}{g\eta_b}\right)
\left(z_c + \frac{1}{g\eta_c}\right)
+\left(z_c + \frac{1}{g\eta_c}\right)
\left(z_a + \frac{1}{g\eta_a}\right)
\right]
\nonumber \\
&=& \eta_a ( 1 + g z_b \eta_b)( 1 + g z_c \eta_c)
+ \eta_b ( 1 + g z_c \eta_c)( 1 + g z_a \eta_a)
+ \eta_c ( 1 + g z_a \eta_a)( 1 + g z_b \eta_b) \, .
\end{eqnarray}
\end{widetext}
is a dimensionless quantity which is proportional to the
total resistance of the circuit (though note that the
total resistance depends on the arrangement of the bias
voltages). The symbols $P,Q,R,S$ stand for
\begin{eqnarray}
P &=& 1+z_a\left(s_{b}^{(c)}+s_{c}^{(c)}\right)
+z_c s_{c}^{(c)} \, ,
\nonumber \\
Q &=& -z_a\left(s_{b}^{(b)}+s_{c}^{(b)}\right)
-z_c s_{c}^{(b)} \, ,
\nonumber \\
R &=& 1+z_a\left(s_{b}^{(b)}+s_{c}^{(b)}\right)
+z_b s_{b}^{(b)} \, ,
\nonumber \\
S &=& -z_a\left(s_{b}^{(c)}+s_{c}^{(c)}\right)
-z_b s_{b}^{(c)} \, .
\label{PQRS}
\end{eqnarray}
Equations \eref{C_1st}--\eref{PQRS} constitute the main results of
this paper; they agree with results obtained from a Langevin
formulation approach.\cite{WY05} The coefficients
$s_{\alpha}^{(\beta)}$, $s_{\alpha\gamma}^{(\beta)}$ here are the
same as those of Eq.~\eref{S_alpha_expand}, whose explicit forms are
provided in Appendix A. Using elementary circuit theory, one can
check easily that $C_\alpha$ ($\alpha=a,b,c$) is exactly the
dimensionless conductance ({\em ie}.~conductance in units of
$e^2/h$) of the arm $A_\alpha$ in the presence of external
impedances. Also, as one can check from the above expressions, when
all external impedances are zero, for example, $C_{b}$ reduces to
the bare moment $(s_{b}^{(b)}+s_{b}^{(c)})=g\eta_a\eta_b$, namely
the conductance of arm $A_b$ (note that we are considering $V_b=V_c$
here, thus the conductance is the sum of the two contributions).

\begin{figure}
\includegraphics*[width=50mm]{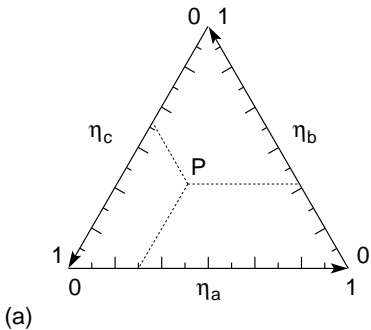}
\caption{\small (a) Illustration for the triangular coordinate
frames adopted for the plots in this paper. For any point $P$ on the
plot, the corresponding $\eta_\alpha$'s can be read off as
indicated. The lower panels plot the {\em bare} cross correlations
(namely, $C_{bc}=s_{bc}$ when $z_a=z_b=z_c=0$) for Y-shaped
conductors with arms of (b) tunnel junctions and (c) diffusive
wires. In (b) and (c) (and also in other plots in this paper) the
unit of $C_{bc}$ is $g$ and the coordinates $x,y$ of the boxed-frame
are related to the $\eta_\alpha$'s by $\eta_a=x$,
$\eta_b=(2/\sqrt{3})y$, and $\eta_c=1-(2/\sqrt{3})y$; the thick
lines over the surfaces depict the contours for $C_{bc}=0$.}
\label{bare_sbc}
\end{figure}

The second moments $C_{bb}$, $C_{cc}$ are the noise correlators and
$C_{bc}$ is the cross correlation between arms $A_b$ and $A_c$
(though see above, below Eq.~\eref{C_coeff}). When all the
$z_\alpha$'s are zero, $C_{bc}$ reduces to the bare cross
correlation $s_{bc}\equiv s_{bc}^{(b)}+s_{bc}^{(c)}$. As is well
known, due to the Fermi statistics of the charge carriers, $s_{bc}$
is always negative.\cite{FS} We plot in Fig.~\ref{bare_sbc} the bare
cross correlations $s_{bc}$ ($C_{bc}$ for $z_a=z_b=z_c=0$) for
Y-shaped conductors made of tunnel junctions and of diffusive wires.
In presenting our results, here and below we shall adopt a
triangular coordinate frame (see Fig.~\ref{bare_sbc}(a)). This is
because for given values of $z_\alpha$'s it is convenient to plot
$C_{bc}$ for all possible values of the $\eta_\alpha$'s using this
system of frames. As one can check easily from
Fig.~\ref{bare_sbc}(a), every point on the triangle satisfies the
constraint $\eta_a+\eta_b+\eta_c=1$.

It should be noticed that, as a consequence of the feedback effects
from voltage fluctuations, the second moments are now linear
combinations of all bare second moments, instead of simple rescaling
of the corresponding bare moments. As demonstrated in Appendix C,
the noise correlators $C_{bb}$, $C_{cc}$ are always positive, as
they should. However, the cross correlation $C_{bc}$ can change sign
in different parameter regimes. This is in sharp contrast with the
bare cross correlation $s_{bc}$, which is always negative. As we
will see, this occurs because the external impedances induce voltage
fluctuations which can revert the sign of the cross correlations. In
the next section, we will study this feedback effect in detail. In
particular, we will be interested in the cases with the cross
correlation $C_{bc}$ turning {\em positive}.

\section{The cross correlation $C_{bc}$}
\label{sec_cross_corrln}
\begin{figure}
\includegraphics*[width=80mm]{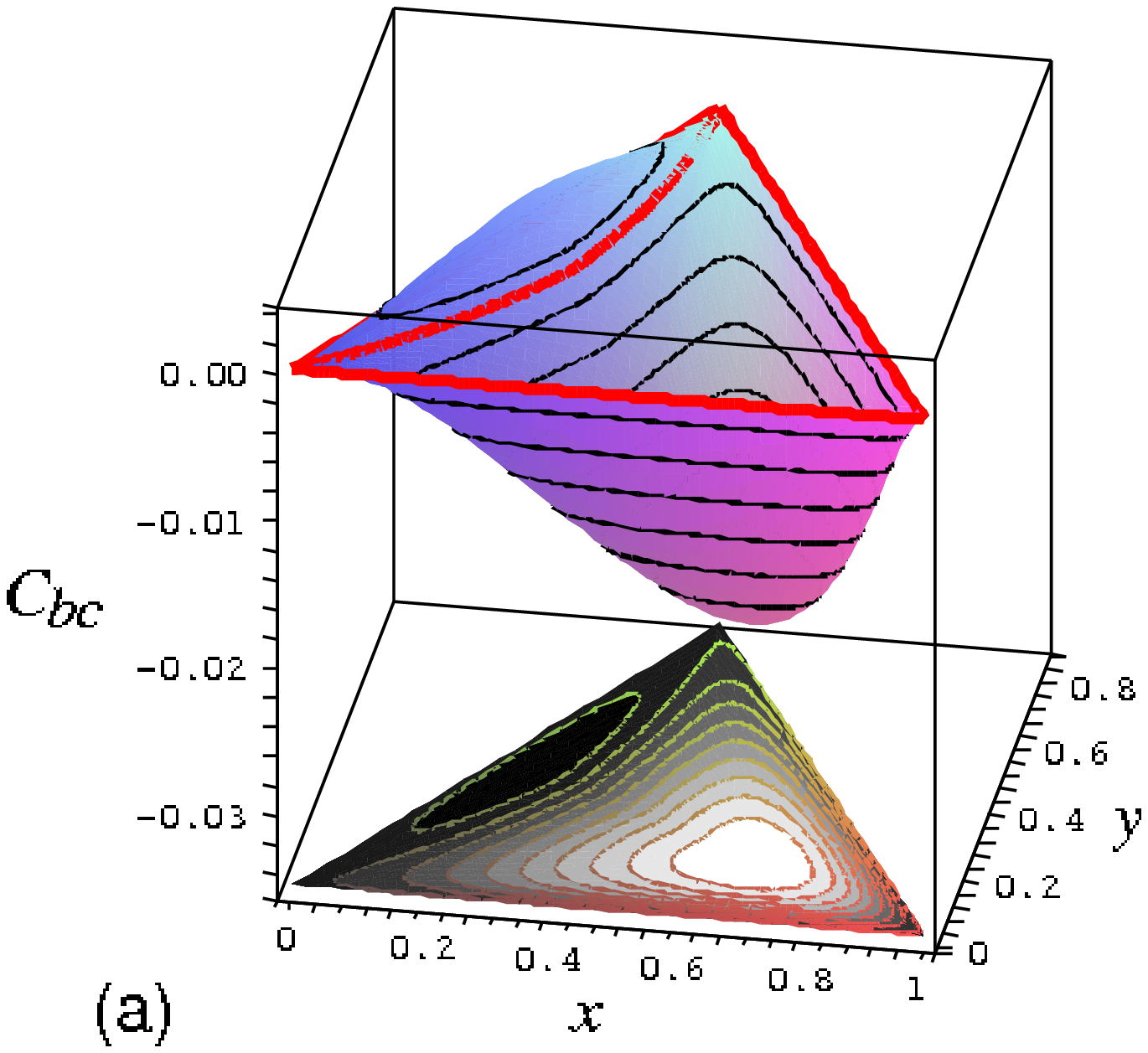}
\includegraphics*[width=80mm]{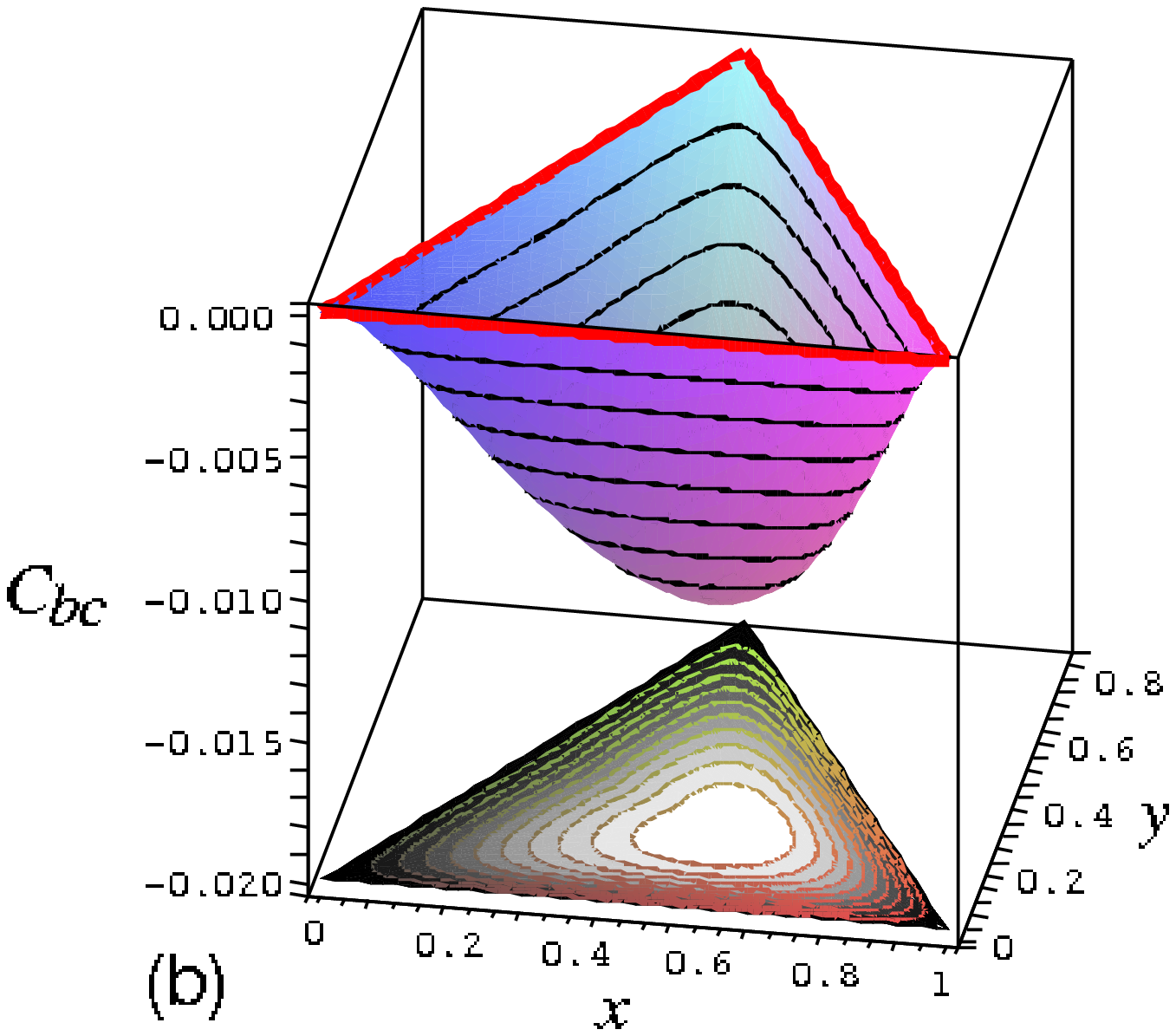}
\caption{\small Plots for the cross correlations for Y-shaped
conductors with arms of (a) tunnel junctions and (b) diffusive wires
where the external impedances are $z_a=z_b=z_c=1/g$.}
\label{zzz_1_3d}
\end{figure}

\begin{figure}
\includegraphics*[width=80mm]{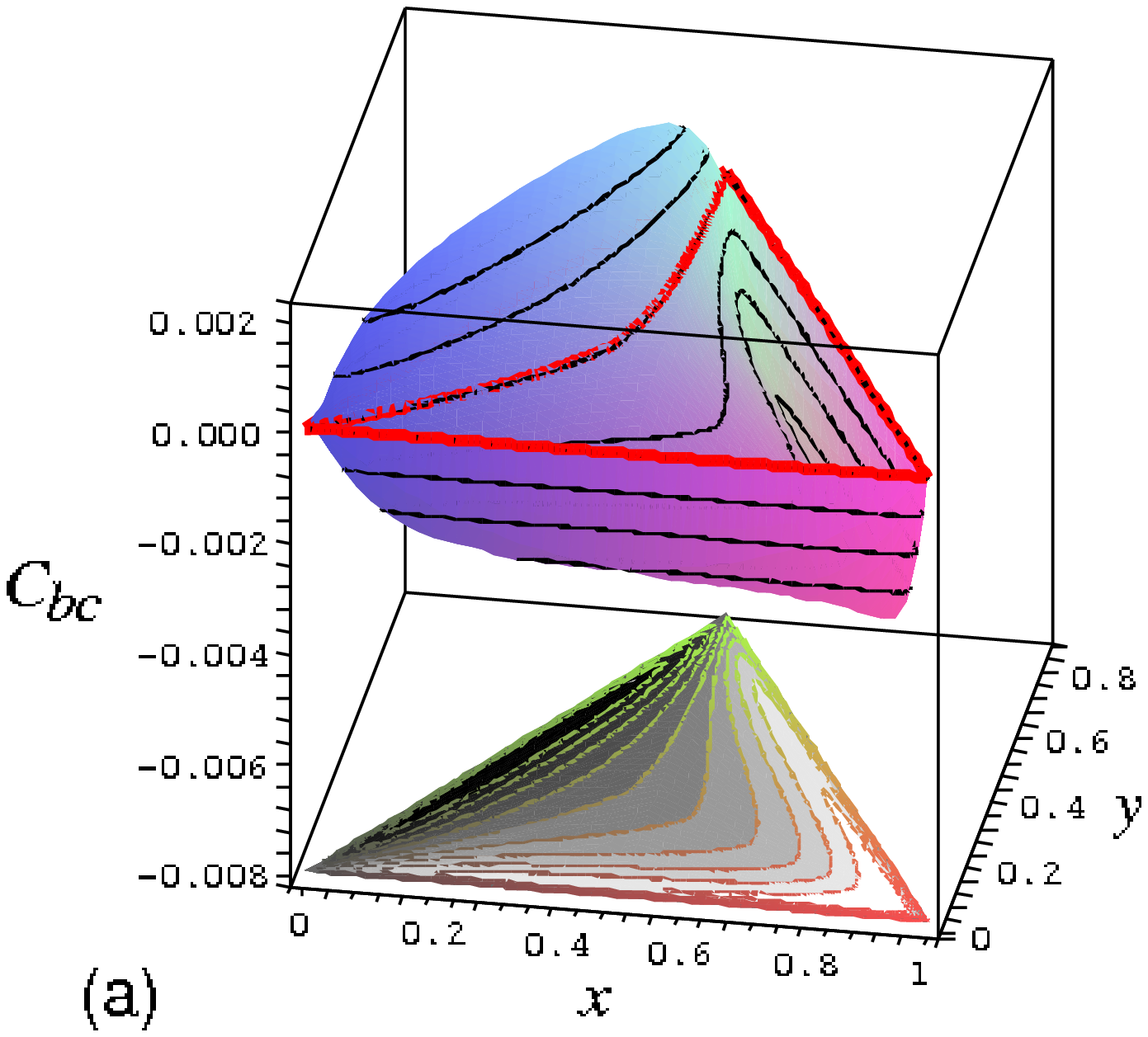}
\includegraphics*[width=80mm]{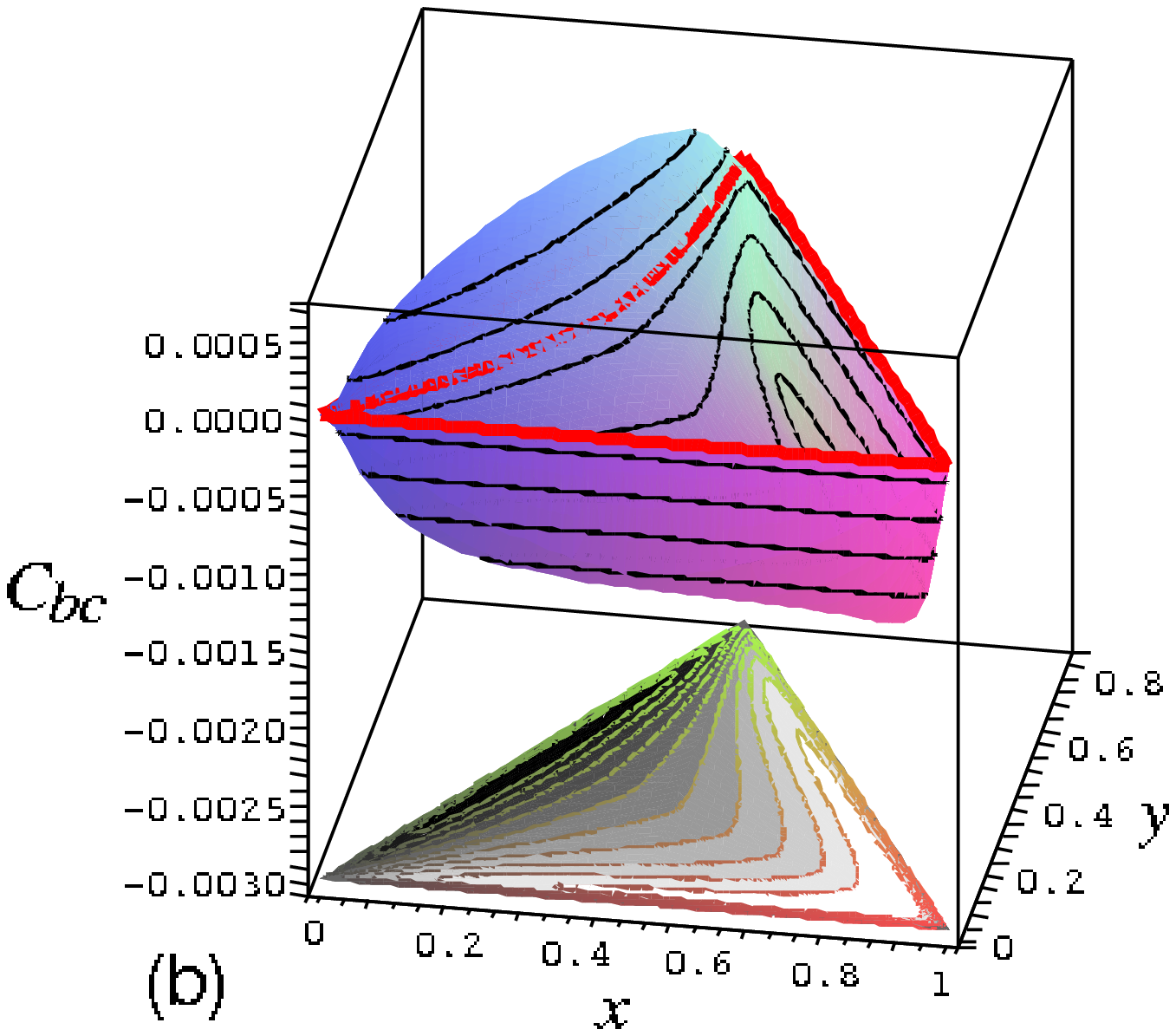}
\caption{\small Plots for the cross correlations of Y-shaped
conductors with (a) tunnel junctions and (b) diffusive wires in the
arms. Here the external impedances are $z_a=z_b=z_c=10g^{-1}$. Note
the different vertical scales in these two plots.} \label{zzz_10_3d}
\end{figure}

%=======================================
In this section, we present our results for the cross correlation
$C_{bc}$ in different parameter regimes. The parameter space we are
exploring consists of six non-negative parameters
$\eta_a,\eta_b,\eta_c,z_a,z_b,z_c$ subjected to the constraint
$\eta_a+\eta_b+\eta_c=1$. As explained in Appendix C, it is useful
here to divide the parameter space into two regions
\begin{eqnarray}
&\mbox{region I} :& z_a\eta_a >
\mbox{$z_b\eta_b$ and $z_c\eta_c$,}
\nonumber \\
&\mbox{region II} :& z_a\eta_a <
\mbox{$z_b\eta_b$, or $z_c\eta_c$, or both.}
\label{regions}
\end{eqnarray}
One can show that for region I the cross correlation $C_{bc}$ is
always negative, while for region II it can flip sign (see Appendix
C). For example, if $z_a\neq 0$ while $z_b=z_c=0$, the cross
correlation must be negative no matter what values the conductances
$\eta_a,\eta_b,\eta_c$ may be. In the following, we will be
interested primarily in region II of the parameter space. Since in
experiments the external impedances are usually of the order of the
sample resistance,\cite{RSP} we shall set $z_\alpha$'s of the order
of $1/g$ and plot the cross correlation $C_{bc}$ for all values of
the $\eta_\alpha$'s. Typical plots for $C_{bc}$ are shown in
Figs.~\ref{zzz_1_3d} and \ref{zzz_10_3d} (see also Figs.~2 and 3 in
Ref.~\onlinecite{WY05}). One can observe clearly regions of positive
cross correlations which grow with increasing external impedances.
We shall now provide physical pictures for these results.

Let us examine first the case $z_a=0$, $z_b=z_c\equiv z\neq 0$. This
corresponds to the situation when one attempts to measure the cross
correlation $C_{bc}$ by connecting the arms $A_b$ and $A_c$ to
voltage probes, while leaving $A_a$ unmeasured. As is obvious from
Eq.~\eref{regions}, in this case we are entirely in region II of the
parameter space since $z_a\eta_a=0$ always. It is therefore possible
to have large areas of positive cross correlations in the
$\eta$-triangle, as discovered in Ref.~\onlinecite{WY05} (see
Figs.~2 and 3 there). For instance, for $z=10g^{-1}$, it was found
that the cross correlation is almost always positive. In fact, the
large $z$ behavior of $C_{bc}$ is determined by the coefficient of
its leading term. It is not hard to check that for $z_a=0$ while
$z_b=z_c\neq 0$, the leading coefficient of $C_{bc}$ is {\em always}
positive. Therefore, for $z$ large enough, $C_{bc}$ can be positive
over the whole $\eta$-triangle ({\em ie.}, whatever values of the
$\eta_\alpha$'s)!

To understand the above results, we note that an interesting feature
in the plots for $C_{bc}$ (see Figs.~2 and 3 in
Ref.~\onlinecite{WY05} and also Figs.~\ref{zzz_1_3d} and
\ref{zzz_10_3d} above) is that positive cross correlations develop
more easily for small $\eta_a$ and large $z_b$, $z_c$. This can be
understood physically as follows (for more quantitative analysis,
see later). When there is a positive fluctuation of the current
through, say the arm $A_b$, there is a corresponding increase in the
potential at node $2$ in Fig.~\ref{circuit}. This voltage
fluctuation in turn will lead to an extra current through the arm
$A_c$, thus giving a {\it positive} contribution to the cross
correlation $C_{bc}$. This contribution will in particular be large
for small $\eta_a$, since most of this fluctuating current will flow
through $c$. We have a net positive $C_{bc}$ if this contribution
overwhelms the ``bare" negative correlation contribution given by
$s_{bc}$. Therefore, the positive region starts near small $\eta_a$,
and grows with increasing $z_b$ and $z_c$.

More quantitatively, let us consider the case with $\eta_b=\eta_c$.
Recalling that $z_a=0$ and $z_b=z_c\equiv z$ here, one gets from
Eq.~\eref{PQRS} for small $\eta$ that $P=R=1+\zeta$ and $Q=S=\zeta$
with $\zeta\equiv gz\eta_b^2$. From \eref{Cbc} we then find
$C_{bc}\propto (1+2\zeta) \left[\zeta(1+\zeta) (s_{bb} + s_{cc} + 2
s_{bc}) + s_{bc} \right]$. Note that $(s_{bb} + s_{cc} + 2 s_{bc})$
is proportional to the bare shot noise for the total current and
hence positive definite. Thus for sufficiently large $z$ (hence
$\zeta$), one can have $C_{bc}>0$. Note also that for tunnel
junctions $s_{bc}$ is proportional to $\eta_a^2$, whereas for
diffusive wires it is proportional to $\eta_a$ (see Appendix A).
This explains why it is easier for tunnel junctions to get positive
cross correlations (see Figs.~2 and 3 in Ref.~\onlinecite{WY05} and
also Figs.~\ref{zzz_1_3d} and \ref{zzz_10_3d} above).

Let us now turn to another case which may also have experimental
interest. Suppose one connects the sample to three voltage probes
with (almost) identical impedances, the circuit then corresponds to
$z_a=z_b=z_c\equiv z$ in our calculation. Our results for $z=1/g$
are shown in Fig.~\ref{zzz_1_3d}. Similar to the previous case, for
tunnel junctions, there is an appreciable region of positive cross
correlation, while none for the diffusive wires (cf.~Fig.~2 in
Ref.~\onlinecite{WY05}). According to our analysis above, for larger
values of $z$ the region for positive cross correlations should
grow. Figure \ref{zzz_10_3d} shows the results for $z=10g^{-1}$; the
trend is clearly in accordance with what we concluded above. We
remark that the region for positive cross correlations is here
bounded by the lines $\eta_a<\eta_b$ and $\eta_a<\eta_c$. As
discussed in Appendix C, for $z_a=z_b=z_c$ one can show that
whenever $\eta_a$ is greater than $\eta_b$ or $\eta_c$, one must
have $C_{bc}<0$. Note that this restricts further the region for
positive cross correlations [cf., region II of \eref{regions}].

Finally, we consider the simple case $z_a=z_c=0$ while $z_b\neq 0$.
Though may not of practical interest, this case may help further
understand our results. In the large $z_b$ limit, one can check that
the leading coefficient of $C_{bc}$ would be positive if
$\eta_a<\eta_c$. Physically, this is because the resistor connected
to the arm $A_b$ causes voltage fluctuations which act back on the
current. The reversed current would go into arm $A_a$ if the
conductance of arm $A_a$ is greater than that of $A_c$ ({\em ie.},
$\eta_a>\eta_c$). However, if $\eta_a<\eta_c$ the fluctuating
current would go into arm $A_c$, leading to enhancement of the
current $I_c$. This is the reason for the positive cross correlation
$C_{bc}$ between arms $A_b$ and $A_c$.

\section{Discussions and conclusions}
\label{sec_conclu}
%====================================
In summary, we have applied a Keldysh technique to study effects of
voltage fluctuations on the current correlations in a mesoscopic,
phase coherent Y-shaped conductor that is connected to a measuring
circuit with finite impedance. We find that, at zero temperature and
at low frequencies, the current moments are significantly modified
by the feedbacks from voltage fluctuations due to the finite
impedance in the circuit. In particular, the current moments cannot
be obtained from simple rescaling of the bare moments already in the
second moments. An interesting consequence of this is that there can
be positive cross correlations in appropriate parameter regimes.

\begin{figure}
\includegraphics*[width=70mm]{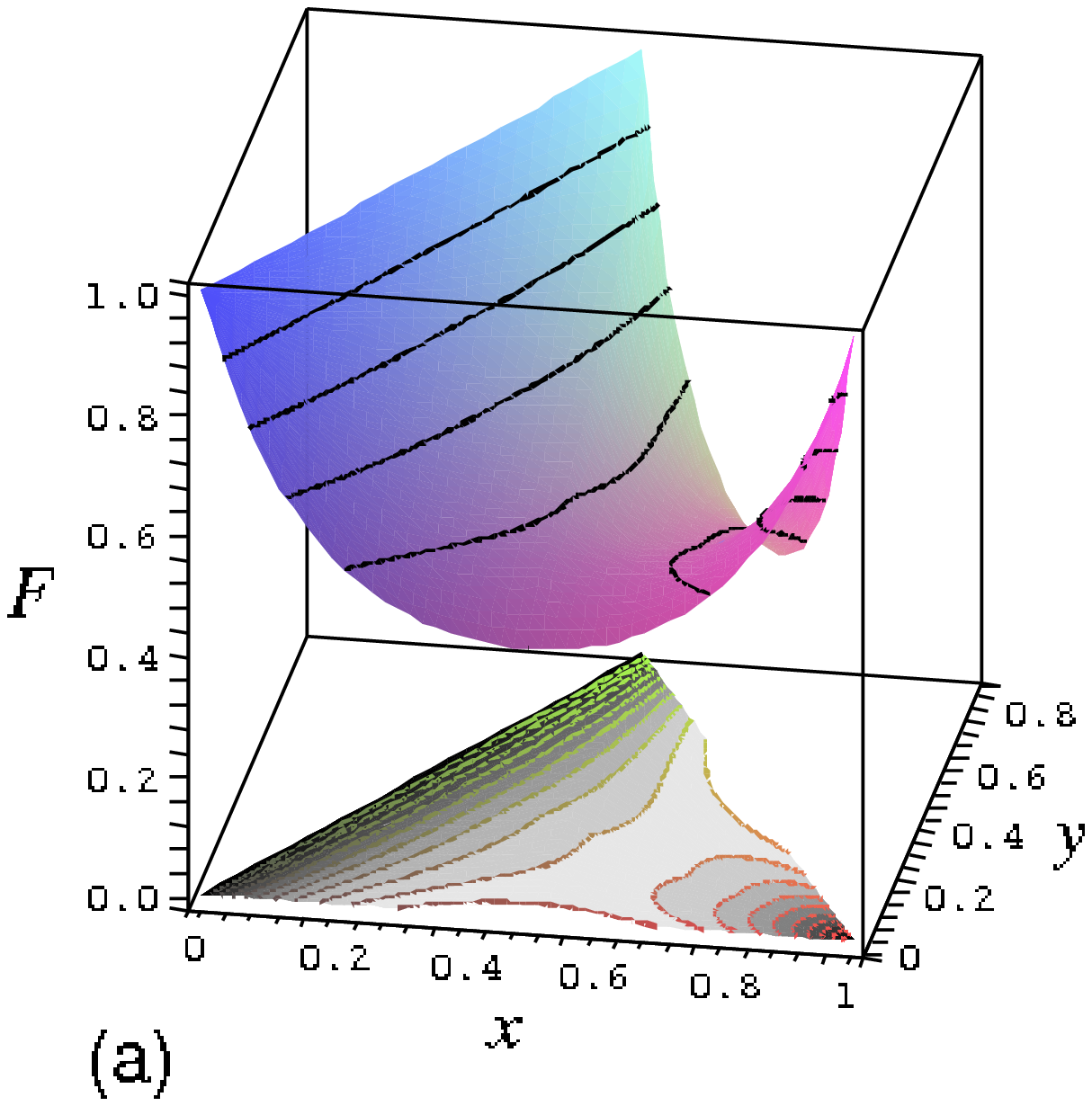}
\includegraphics*[width=70mm]{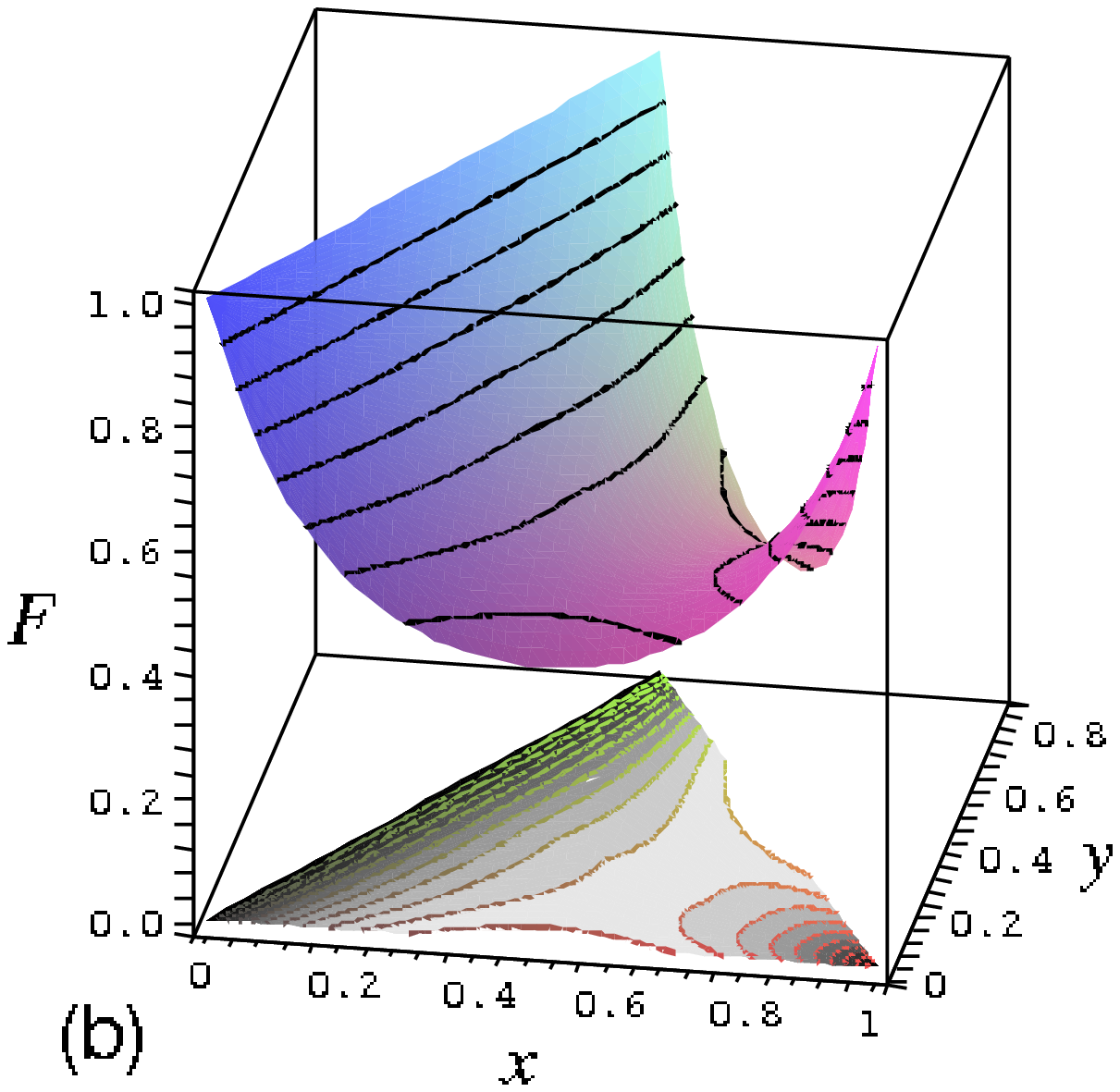}
\caption{\small The Fano factors for Y-shaped conductors with (a)
tunnel junctions and (b) diffusive wires in the arms. Here the
impedances are $z_a=z_b=z_c=1/g$ and we have normalized the Fano
factors to their bare values.} \label{Fano}
\end{figure}

Positive cross correlations are usually associated with ``bunching"
behavior of charge carriers, which would also imply enhanced Fano
factors. However, it is also known that positive cross correlation
is not a {\em sufficient} condition for electron bunching. Indeed
there are existing examples of positive cross correlations that do
not go with electron bunching.\cite{CBB} We have calculated the Fano
factors for the total current in our circuit and find that there is
in fact no any electron bunching here -- even when the cross
correlations are positive. For example, for $z_a=z_b=z_c=1/g$ we
plot in Fig.~\ref{Fano} the Fano factors\cite{notes_Fano} $F$ for
the total current for tunnel junctions and diffusive wires
normalized to their bare values. It is seen that the Fano factors at
most recover their bare values in the regions with positive cross
correlations and never go beyond them. Thus there is no any trace of
electron bunching here. The positive cross correlation we have found
is therefore, instead of electron bunching, purely a feedback effect
due to voltage fluctuations. Our result thus shows that one has to
be very careful in interpreting experimental data that exhibit
positive cross correlations, especially when the measuring circuit
is expected to have a finite impedance.

\begin{acknowledgments}
The authors would like to thank Professor Markus B\"uttiker for
comments. This research was supported by NSC of Taiwan under grant
numbers NSC93-2112-M-001-016 for SKY and NSC93-2112-M-194-019,
NSC94-2112-M-194-008 for STW. We would also like to thank NCTS
Hsinchu Taiwan for support in part of this work.
\end{acknowledgments}

\appendix
%========
\section{Bare current moments and the $s$-coefficients}
%======================================================
In the Keldysh technique, current moments of any order
can be obtained from derivatives of the generation
functional with respect to the auxiliary field
$\sigma$.\cite{KNB,Kind} Therefore, to obtain current
moments up to second order, we expand the action for
the sample to quadratic order
\begin{eqnarray}
\tau S_A &=&
\langle \hat{q}_b \rangle (\sigma_1-\sigma_2) +
\langle \hat{q}_c \rangle (\sigma_1-\sigma_3) +
\nonumber \\ &&
\langle\langle \hat{q}_b^2 \rangle\rangle
(\sigma_1-\sigma_2)^2/2! +
\langle\langle \hat{q}_c^2 \rangle\rangle
(\sigma_1-\sigma_3)^2/2! +
\nonumber \\ &&
\langle\langle \hat{q}_b \hat{q}_c \rangle\rangle
(\sigma_1-\sigma_2)(\sigma_1-\sigma_3) \, .
\end{eqnarray}
Here $\hat{q}_\alpha=(1/e)\int_0^\tau \hat{I}_\alpha dt$ is the
operator for transferred charges through arm $A_\alpha$ during the
time interval $\tau$; $\langle\,\cdots\rangle$ is a quantum average
over ensembles and $\langle\langle\,\cdots\rangle\rangle$ is the
irreducible (or cumulant) average:
$\langle\langle\hat{P}\hat{Q}\rangle\rangle\equiv\langle\hat{P}\hat{Q}\rangle
- \langle\hat{P}\rangle\langle\hat{Q}\rangle$.

As illustrated in Fig.~\ref{circuit}, the arms of sample $A$ are
connected to nodes with (intermediate) phase variables $\phi_1$,
$\phi_2$, $\phi_3$. Since all external impedances are beyond this
region, the related moments $\langle\hat{q}_\alpha\rangle$,
$\langle\langle\hat{q}_\alpha\hat{q}_\beta\rangle\rangle$ are thus
simply the {\em bare} ones. These moments of transferred charges are
related to the $s$-coefficients (which are proportional to the {\em
bare} current moments) defined in the text. Comparing the above
equation with Eqs.~\eref{S_A} and \eref{S_alpha_expand}, we find the
first moments
\begin{eqnarray}
\langle \hat{q}_\alpha \rangle = s_{\alpha}^{(b)} (\phi_1-\phi_2) +
s_{\alpha}^{(c)} (\phi_1-\phi_3)
\end{eqnarray}
and the second moments
\begin{eqnarray}
\langle\langle \hat{q}_\alpha \hat{q}_\beta \rangle\rangle =
s_{\alpha\beta}^{(b)}(\phi_1-\phi_2) +
s_{\alpha\beta}^{(c)}(\phi_1-\phi_3) \, .
\end{eqnarray}
The $s$-coefficients can be calculated using the technique developed
by Nazarov and coworkers\cite{Nazarov} and also in other
methods.\cite{Yip05} To avoid distractions from the main points, we
shall not present these calculations.\cite{WY_unpub} We shall simply
list the results and point out the essential properties (such as
their sign characteristics) which will be crucial to our
calculation.

\subsection{Tunnel junctions}
%----------------------------
When the arms of sample $A$ are made of tunnel junctions, we find
the first moments yield
\begin{eqnarray}
s_{b}^{(b)} =  g \eta_b (\eta_a+\eta_c)\, ,
\qquad
s_{b}^{(c)} = -g \eta_b \eta_c \, ,
\nonumber \\
s_{c}^{(b)} = -g \eta_b \eta_c \, ,
\qquad
s_{c}^{(c)} =  g \eta_c (\eta_a+\eta_b)\, .
\label{s_1}
\end{eqnarray}
The second moments take different forms depending on the relative
magnitudes of the intermediate phase variables $\phi_1$, $\phi_2$,
and $\phi_3$. For $\phi_1>\phi_2>\phi_3$, one can find
\begin{eqnarray}
s_{bb}^{(b)} &=&
g \eta_b [ \eta_a (1-2\eta_a\eta_b)
- \eta_c (1-2\eta_b\eta_c)] \, ,
\nonumber\\
s_{bb}^{(c)} &=&
g \eta_b\eta_c (1-2\eta_b\eta_c) \, ,
\nonumber\\
s_{cc}^{(b)} &=&
-g \eta_b\eta_c(1-2\eta_a\eta_c-2\eta_b\eta_c) \, ,
\nonumber\\
s_{cc}^{(c)} &=&
g \eta_c [ \eta_a (1-2\eta_a\eta_c)
+ \eta_b (1-2\eta_a\eta_c-2\eta_b\eta_c) ]\, ,
\nonumber\\
s_{bc}^{(b)} &=&
g \eta_b\eta_c(1-2\eta_a^2-2\eta_a\eta_c-2\eta_b\eta_c) \, ,
\nonumber\\
s_{bc}^{(c)} &=&
- g \eta_b\eta_c(1-2\eta_a\eta_c-2\eta_b\eta_c) \, .
\label{ss_tunl}
\end{eqnarray}
For $\phi_1>\phi_3>\phi_2$,\cite{notes_phi} base on simple symmetry
considerations, one can obtain the expressions for the
$s$-coefficients by exchanging the indices $b$, $c$ in the formulas
above. For example, $s_{cc}^{(b)}$ can be obtained from the above
expression for $s_{bb}^{(c)}$ with all indices of its right hand
members making the exchange $b\leftrightarrow c$.

From these expressions for the $s$-coefficients, it is not difficult
to show that when $\phi_1>\phi_2>\phi_3$ the second order
coefficients satisfy
\begin{eqnarray}
s_{bb}^{(c)} > 0 , \quad s_{cc}^{(c)} > 0,
\quad s_{cc}^{(b)} < 0, \quad s_{bc}^{(c)} < 0
\label{s_signs_parts}
\end{eqnarray}
always, while $s_{bb}^{(b)}$ and $s_{bc}^{(b)}$ do not have definite
signs for general values of the $\eta_\alpha$'s. For example, using
$\eta_a+\eta_b+\eta_c=1$, one can write from \eref{ss_tunl}
\begin{eqnarray}
s_{bb}^{(c)} &=&
g \eta_b\eta_c [(\eta_a+\eta_b+\eta_c)^2-2\eta_b\eta_c] \, ,
\end{eqnarray}
which is obviously always positive since all $\eta_\alpha$'s are
non-negative. Similarly one can check for the other coefficients. We
shall make use of these properties in the following when determining
the signs of the renormalized (current) moments (the
$C$-coefficients). In the case of $\phi_1>\phi_3>\phi_2$, as pointed
out above, everything follows from exchanging $b$ and $c$.
Therefore, it is now the $s^{(b)}_{\alpha\beta}$ terms that have
definite sign, while not the $s^{(c)}_{\alpha\beta}$ terms (except
$s^{(c)}_{bb}$, though).

In our calculation we also find it convenient to introduce the
``combined" $s$-coefficients
\begin{eqnarray}
s_{\alpha} \equiv s_{\alpha}^{(b)} + s_{\alpha}^{(c)}
\quad \mbox{and} \quad
s_{\alpha\beta} \equiv
s_{\alpha\beta}^{(b)} + s_{\alpha\beta}^{(c)} .
\label{s_sum}
\end{eqnarray}
Therefore, we have
\begin{eqnarray}
s_{b} &=&  g \eta_a \eta_b\, ,
\quad
s_{c} =  g \eta_a \eta_c\, ,
\quad
s_{bc} =
-2 g \eta_a^2\eta_b\eta_c \, ,
\nonumber \\
s_{bb} &=&
g \eta_a \eta_b (1-2\eta_a\eta_b) \, ,
\quad
s_{cc} =
g \eta_a \eta_c (1-2\eta_a\eta_c) \, .
\nonumber \\ &&
\label{s_sum_tunl}
\end{eqnarray}
It is easy to check that for the second order terms
\begin{eqnarray}
s_{bb} > 0 , \quad  s_{cc} > 0, \quad s_{bc} < 0 \, .
\label{s_signs}
\end{eqnarray}
These are indeed what one would have expected, in view of the fact
that these $s$-coefficients are exactly the {\em bare} current
moments (namely the current moments at zero impedance -- see text):
$s_{bb}$, $s_{cc}$ are the {\em bare} noise correlators and $s_{bc}$
the {\em bare} cross correlations, which must take the signs above.
Note that Eqs.~\eref{s_sum_tunl} and \eref{s_signs} are valid for
both possible arrangements of the intermediate phase variables.

\subsection{Diffusive wires}
%------------------------------
When the sample arms are made of diffusive wires, calculation shows
that the first order coefficients are the same as those for tunnel
junctions, namely Eq.~\eref{s_1}. In the case of
$\phi_1>\phi_2>\phi_3$, we find the second order coefficients
\begin{eqnarray}
s_{bb}^{(b)} &=&
\frac{g}{3} \eta_b (\eta_a-\eta_c) \, ,
\qquad
s_{bb}^{(c)} =
\frac{g}{3} \eta_b\eta_c (1+2\eta_a) \, ,
\nonumber \\
s_{cc}^{(b)} &=&
\frac{g}{3} \eta_b\eta_c (2\eta_a-1) \, ,
\qquad
s_{cc}^{(c)} =
\frac{g}{3} \eta_c (\eta_a+\eta_b) \, ,
\nonumber\\
s_{bc}^{(b)} &=&
\frac{g}{3} \eta_b\eta_c (1-2\eta_a) \, ,
\qquad
s_{bc}^{(c)} =
- \frac{g}{3} \eta_b\eta_c \, .
\end{eqnarray}
As before, for $\phi_1>\phi_3>\phi_2$ one can obtain the expressions
for the $s$-coefficients by exchanging $b$ and $c$ in the formulas
above.

Just like for tunnel junctions, here we also define the ``combined"
$s$-coefficients as in Eq.~\eref{s_sum}. Thus, we have $s_b$, $s_c$
the same as in Eq.~\eref{s_sum_tunl} and
\begin{eqnarray}
s_{bb} &=&
\frac{g}{3} \eta_a\eta_b (1+2\eta_c) \, ,
\qquad
s_{cc} =
\frac{g}{3} \eta_a\eta_c (1+2\eta_b) \, ,
\nonumber \\
s_{bc} &=&
- \frac{2g}{3} \eta_a\eta_b\eta_c \, .
\end{eqnarray}
As one can check easily, the $s$-coefficients for diffusive wires
also have the same sign characteristics as those for tunnel
junctions. In other words, Eqs.~\eref{s_signs_parts} and
\eref{s_signs} are valid here as well.

\section{Algebra for solving the $C$-coefficients}
%=================================================
In this Appendix, we summarize briefly the way to solve for the
$C$-coefficients.

Although in calculating the renormalized current moments one shall
need only the expansion of $\sigma_1$ (see Eq.~\eref{tau_S_tot}), to
find the expansion coefficients, however, one has to solve the
coupled saddle point equations \eref{saddle}. Therefore, we expand
all the intermediate variables $\sigma_k$, $k=1,2,3$ as the
following
\begin{eqnarray}
\frac{(\sigma_{\alpha_k}-\sigma_k)}{z_{\alpha_k}} &\simeq&
C_{b}^{(k)} (\sigma_a-\sigma_b) +
C_{c}^{(k)} (\sigma_a-\sigma_c)
\nonumber \\ &&
+ \frac{C_{bb}^{(k)}}{2!} (\sigma_a-\sigma_b)^2
+ \frac{C_{cc}^{(k)}}{2!} (\sigma_a-\sigma_c)^2
\nonumber \\ &&
+ C_{bc}^{(k)} (\sigma_a-\sigma_b)(\sigma_a-\sigma_c) \, ,
\end{eqnarray}
where the subscript
\begin{eqnarray}
\alpha_k = \left\{
    \begin{array}{cc}
        a & \mbox{for $k=1$,}\\
        b & \mbox{for $k=2$,}\\
        c & \mbox{for $k=3$.}
    \end{array}
           \right.
\end{eqnarray}
Substituting the above expansions into the saddle point
equations \eref{saddle} and comparing coefficients, we get the
simultaneous equations for the $C$-coefficients, which can then
be solved easily. Since we shall need only the $C^{(1)}$
coefficients for our calculation, the superscripts are omitted
in the text.

\section{Sign characteristics of the renormalized second order moments}
%======================================================================
In this Appendix we examine the sign characteristics of the
renormalized second order moments (the second order
$C$-coefficients). We will show that the noise correlators $C_{bb}$,
$C_{cc}$ are always positive, while the cross correlation $C_{bc}$
can change sign in different parameter regimes.

We study first the noise correlator $C_{bb}$ and consider the
case $\phi_1>\phi_2>\phi_3$. As noted above, in this case
$s_{bb},s_{cc}$ and $s_{bb}^{(c)},s_{cc}^{(c)}$ are always
positive, while $s_{bc},s_{bc}^{(c)}$ are always negative (see
Eqs.~\eref{s_signs_parts}, \eref{s_signs}).
To determine the sign of $C_{bb}$, it is therefore preferable
to eliminate the $s^{(b)}$ coefficients in favor of the
$s^{(c)}$'s. Noting that
\begin{eqnarray}
(Q+R) = (P+S) + g\eta_a (z_b\eta_b-z_c\eta_c) \, ,
\label{QR_PS}
\end{eqnarray}
one can rewrite the terms in the braces in Eq.~\eref{Cbb} as
\begin{eqnarray}
\Gamma_{bb} &\equiv&
(P+S) \left[ P^2 s_{bb} + Q^2 s_{cc} + 2PQ s_{bc} \right]
\nonumber \\ &&
+ \,\, g\eta_a (z_b\eta_b-z_c\eta_c)
\nonumber \\ &&
\times
\left[P^2 s_{bb}^{(c)}+Q^2 s_{cc}^{(c)}+2PQ s_{bc}^{(c)}\right]
\, .
\label{temp_b2}
\end{eqnarray}
From the explicit forms of the $s$-coefficients, one can show
that the following inequalities hold
\begin{eqnarray}
(s_{bb}) (s_{cc}) \geq (s_{bc})^2 ,\quad
(s_{bb}^{(c)}) (s_{cc}^{(c)}) \geq (s_{bc}^{(c)})^2 \, .
\end{eqnarray}
Therefore, one has
\begin{eqnarray}
&& (P^2 s_{bb} + Q^2 s_{cc} + 2PQ s_{bc})
\nonumber \\
&\geq& 2 |PQ| \sqrt{(s_{bb})(s_{cc})} + 2PQ s_{bc}
\nonumber \\
&\geq& 2 |PQ| |s_{bc}| + 2 PQ s_{bc} \geq 0 \, .
\end{eqnarray}
Similarly, one can show that $(P^2 s_{bb}^{(c)}+Q^2 s_{cc}^{(c)}+2PQ
s_{bc}^{(c)})\geq 0$ also. For the prefactors in front of the two
terms: from the explicit forms of the $s$-coefficients, obviously
$(P+S)\geq 0$; moreover, it is easy to show that for
$\phi_2>\phi_3$, one must have $z_b\eta_b \geq z_c\eta_c$.
Therefore, we conclude that the quantity in Eq.~\eref{temp_b2} is
always positive and hence $C_{bb}$.

For the case with $\phi_3>\phi_2$, the proof proceeds very
similarly. However, in this case, it is the coefficients
$s_{bb}^{(b)},s_{cc}^{(b)}$ that are always positive and
$s_{bc}^{(b)}$ always negative. It is therefore helpful to express
$(P+S)$ in terms of $(Q+R)$, eliminating the $s^{(c)}$'s in favor of
the $s^{(b)}$'s. Using
\begin{eqnarray}
(P+S) = (Q+R) + g\eta_a (z_c\eta_c-z_b\eta_b) \,
\end{eqnarray}
one can then proceed as before. Noting that $(Q+R)>0$ always and
that when $\phi_3>\phi_2$ one must have $z_c\eta_c>z_b\eta_b$, one
can show easily that $C_{bb}$ is again positive definite. Similar
calculations as above (with $P,Q$ replaced by $S,R$) can also show
that $C_{cc}$ is always positive. Indeed, these results can be
anticipated as $C_{bb}$ and $C_{cc}$ are both autocorrelators and
hence must be always non-negative.

We now turn to the cross correlation $C_{bc}$ and demonstrate the
calculation for the case $\phi_1>\phi_2>\phi_3$; the calculation for
$\phi_1>\phi_3>\phi_2$ proceeds very similarly. As for $C_{bb}$, we
use Eq.~\eref{QR_PS} and rewrite the braced terms in Eq.~\eref{Cbc}
as
\begin{eqnarray}
\Gamma_{bc}&\equiv&
(P+S) \left[PS s_{bb}+QR s_{cc}+(PR+QS) s_{bc}\right]
\nonumber \\ &&
+ \, \, g\eta_a (z_b\eta_b-z_c\eta_c)
\nonumber \\ &&
\times
\left[PS s_{bb}^{(c)}+QR s_{cc}^{(c)}+(PR+QS) s_{bc}^{(c)}\right]
\, .
\label{Gamma_bc}
\end{eqnarray}
Note again that $z_b\eta_b>z_c\eta_c$ here (as $\phi_2>\phi_3$) and
that $(P+S)$, $s_{bb}$, $s_{cc}$, $s_{bb}^{(c)}$, and $s_{cc}^{(c)}$
are always positive, while $s_{bc}$ and $s_{bc}^{(c)}$ are always
negative. Applying the explicit forms of $Q$ and $S$, one can show
easily that once $z_a\eta_a$ is greater than both $z_b\eta_b$ and
$z_c\eta_c$, the quantity $\Gamma_{bc}$ will be negative definite,
hence follows the negative cross correlation. However, if
$z_a\eta_a$ is less than $z_b\eta_b$, or $z_c\eta_c$, or both,
$\Gamma_{bc}$ can take negative or positive values, and thus {\em
positive} cross correlations. This motivates us to divide the
parameter space (consisting of $\eta_a,\eta_b,\eta_c,z_a,z_b,z_c$)
into two parts according to the sign characteristics of $C_{bc}$: in
one part (region I) it is always negative, while in the other
(region II) it can change sign (see Eq.~\eref{regions}).

For the special case with $z_a=z_b=z_c\equiv z$, we note that we can
further limit the region of positive cross correlations to the area
with $\eta_a$ less than {\em both} $\eta_b$ and $\eta_c$. This can
be proved straightforwardly (though the algebra is tedious) using
the explicit forms of $C_{bc}$ and the $s$-coefficients. By
expressing $C_{bc}$ as a polynomial of $z$ and checking the signs of
the coefficients order by order, one can show that $C_{bc}$ must be
negative provided $\eta_a>\min\{\eta_b,\eta_c\}$. Thus, together
with the general criterion above, one obtains the limits for
positive cross correlations here to be $\eta_a$ less than {\em both}
$\eta_b$ and $\eta_c$.


\begin{thebibliography}{99}
%==========================
\bibitem{BB}
Ya. M. Blanter and M. B\"{u}ttiker,
Phys. Rep. {\bf 336}, 1 (2000).

\bibitem{BS}
C. Beenakker and C. Sch\"{o}nenberger,
Phys. Today {\bf 56}, 37 (2003).

\bibitem{Nazarov_rev}
{\em Quantum Noise in Mesoscopic Physics},
edited by Yu. V. Nazarov (Kluwer, Dordrecht, 2003).

\bibitem{FQHE}
See, for example,
E. Comforti, Y. C. Chung, M. Heiblum, V. Umansky, and D. Mahalu,
Nature {\bf 416}, 515 (2002);
Y. C. Chung, M. Heiblum, and V. Umansky,
Phys. Rev. Lett {\bf 91}, 216804 (2003) and references therein.

\bibitem{QO}
See, for example,
M. O. Scully and M. S. Zubairy,
{\em Quantum Optics}
(Cambridge University Press, Cambridge, 1997).

\bibitem{HBT1}
M. Henny, S. Oberholzer, S. Strunk, T. Heinzel, K. Ensslin,
M. Holland, and C. Schonenberger,
Science {\bf 284}, 296 (1999);
W. D. Oliver, J. Kim, R. C. Liu, and Y. Yamamoto,
Science {\bf 284}, 299 (1999).

\bibitem{HBT2}
H. Kiesel, A. Renz, and F. Hasselbach,
Nature {\bf 418}, 392 (2002).

\bibitem{Goldman}
A recent experimental evidence for the fractional statistics of
quantum Hall quasiparticles is provided by
F. E. Camino, W. Zhou, and V. J. Goldman,
Phys. Rev. B {\bf 72}, 075342 (2005).

\bibitem{Safi}
I. Safi, P. Devillard, and T. Martin,
Phys. Rev. Lett. {\bf 86}, 4628 (2001).

\bibitem{Vish}
S. Vishveshwara,
Phys. Rev. Lett. {\bf 91}, 196803 (2003).

\bibitem{Fradkin}
E.-A. Kim, M. Lawler, S. Vishveshwara, and E. Fradkin,
Phys. Rev. Lett. {\bf 95}, 176402 (2005).

\bibitem{RSP}
B. Reulet, J. Senzier, and D. E. Prober,
Phys. Rev. Lett {\bf 91}, 196601 (2003).

\bibitem{KNB}
M. Kindermann, Yu. V. Nazarov,
and C. W. J. Beenakker,
Phys. Rev. Lett. {\bf 90}, 246805 (2003).

\bibitem{Kind}
M. Kindermann, PhD thesis,
Leiden University (2003).

\bibitem{BKN}
C. W. J. Beenakker, M. Kindermann, and Yu. V. Nazarov,
Phys. Rev. Lett. {\bf 90}, 176802 (2003).

\bibitem{WY05}
S.-T. Wu and S.-K. Yip,
Phy. Rev. B {\bf 72}, 153101 (2005).

\bibitem{RB}
V. Rychkov and M. B\"uttiker, cond-mat/0512534.

\bibitem{FS}
M. B\"uttiker, Phys. Rev. Lett. {\bf 65}, 2901 (1990) and Phys. Rev.
B {\bf 46}, 12485 (1992); Th. Martin and R. Landauer, Phys. Rev. B
{\bf 45}, 1742 (1992); E. V. Sukhorukov and D. Loss, Phys. Rev. B
{\bf 59}, 13054 (1999).

\bibitem{Buttiker_rev}
See, eg., M. B\"uttiker in
Ref.~\onlinecite{Nazarov_rev} (2003).

\bibitem{SC}
M. P. Anantram and S. Datta,
Phys. Rev. B {\bf 53}, 16390 (1996);
J. Torr\'es and Th. Martin,
Eur. Phys. J. B {\bf 12}, 319 (1999);
J. B\"orlin, W. Belzig, and C. Bruder,
Phys. Rev. Lett. {\bf 88}, 197001 (2002);
P. Samuelsson and M. B\"uttiker, Phys.
Rev. Lett. {\bf 89}, 046601 (2002).

\bibitem{Texier}
C. Texier and M. B\"uttiker,
Phys. Rev. B {\bf 62}, 7454 (2000).

\bibitem{Crepieux}
A. Cr\'{e}pieux, R. Guyon, P. Devillard, and T. Martin,
Phys. Rev. B {\bf 67}, 205408 (2003).

\bibitem{Taddei}
F. Taddei and R. Fazio,
Phy. Rev. B {\bf 65}, 134522 (2002).

\bibitem{CBB}
A. Cottet, W. Belzig, and C. Bruder,
Phys. Rev. Lett. {\bf 92}, 206801 (2004).

\bibitem{MB00}
A. M. Martin and M. B\"uttiker,
Phys. Rev. Lett. {\bf 84}, 3386 (2000).

\bibitem{FCS}
L. S. Levitov and G. B. Lesovik, JETP Lett. {\bf 58}, 230 (1993); L.
S. Levitov and G. B. Lesovik, cond-mat/9401004; L. S. Levitov, H.
Lee, and G. B. Lesovik, J. Math. Phys. (N. Y.) {\bf 37}, 4845
(1996).

\bibitem{notes_Fano}
The Fano factor $F$ plotted in Fig.~\ref{Fano} is, in the notations
of Eqs.~\eref{C_1st}--\eref{Cbc}, $C_{aa}/(2C_a)$ devided by the
corresponding expression for $z_a=z_b=z_c=0$ (namely, the bare
value); here $C_{aa} = C_{bb}+C_{cc}+2C_{bc}$ and $C_a=C_b+C_c$.

\bibitem{Nazarov}
Yu. V. Nazarov,
Ann. Phys. (Leipzig) {\bf 8}, SI-193 (1999);
Yu. V. Nazarov and D. A. Bagrets,
Phys. Rev. Lett. {\bf 88}, 196801 (2002).

\bibitem{Yip05}
S.-K. Yip,
Phys. Rev. B {\bf 71}, 085319 (2005).

\bibitem{WY_unpub}
S.-T. Wu and S.-K. Yip, unpublished.

\bibitem{notes_phi}
It can be shown easily that for $V_a>V_b>V_c$,
the intermediate phase variables can only be either
$\phi_1>\phi_2>\phi_3$ or $\phi_1>\phi_3>\phi_2$.

\end{thebibliography}
\end{document}